Intégration de contraintes cinématiques pour le calcul de l'orientation optimisée de l'axe de l'outil en usinage 5 axes

Optimisation of tool axis orientation in 5 axis machining taking into account kinematical constraints

Sylvain LAVERNHE

Docteur

LURPA – ENS de Cachan – Université Paris Sud 11

61 Avenue du Président Wilson, 94235 Cachan Cedex

tél. : 01 47 40 27 65 – fax : 01 47 40 22 20 – lavernhe@lurpa.ens-cachan.fr

Christophe TOURNIER

Maître de conférences

LURPA – ENS de Cachan – Université Paris Sud 11

61 Avenue du Président Wilson, 94235 Cachan Cedex

tél. : 01 47 40 29 96 – fax : 01 47 40 22 20 – tournier@lurpa.ens-cachan.fr

Auteur correspondant

Claire LARTIGUE

Professeur des universités

IUT de Cachan – Université Paris Sud 11 – LURPA – ENS de Cachan

9 Avenue de la division Leclerc, 94234 Cachan Cedex

tél. : 01 47 40 29 86 – fax : 01 47 40 22 20 – lartigue@lurpa.ens-cachan.fr

Intégration de contraintes cinématiques pour le calcul de l'orientation optimisée de l'axe de l'outil en usinage 5 axes

Après avoir détaillé les principales difficultés liées à l'usinage 5 axes UGV, nous présentons un modèle de représentation des trajectoires 5 axes sous forme surfacique permettant de prendre en compte des contraintes géométriques et cinématiques. Ce modèle est intégré dans d'optimisation des trajets 5 axes afin de maximiser la productivité tout en garantissant la qualité géométrique attendue. Un cas d'application est détaillé, illustrant la modification de l'orientation de l'axe de l'outil afin d'améliorer le comportement cinématique des axes lors du suivi des trajectoires.

**fraisage 5 axes/UGV/modèle surfacique de trajectoire/optimisation de trajectoire**

Optimisation of tool axis orientation in 5 axis machining taking into account kinematical constraints

After presenting the main difficulties related to machining 5 axes machining within the context of HSM, we present a surface model of 5-axis trajectories allowing taking into account geometrical and kinematical constraints. This model is integrated in an optimization scheme in order to maximize the productivity while ensuring the expected geometrical quality. A case of application is detailed, illustrating the modification of the tool axis in order to improve axis kinematical behavior during the tool path follow-up.

**5-axis milling/High Speed Machining/tool path surface modelling/tool path optimisation**

$a$ : accélération courante d'axe .................................................................. m.s$^{-2}$

$A$ : accélération maximale d'axe ................................................................. m.s$^{-2}$

$A,C$ : positions des axes de rotation dans le repère associé à la machine ............. rad

$D$ : distance entre passes définie dans l'espace cartésien ....................................... m

$I,J,K$ : cosinus directeur de l'orientation de l'axe de l'outil dans le repère pièce

$j$ : jerk courant d'axe ................................................................................. m.s$^{-3}$

$J$ : jerk maximal d'axe ............................................................................... m.s$^{-3}$

$L$ : longueur de la trajectoire ......................................................................... m

$p$ : position courante d'axe ........................................................................... m

$p_0$ : position initiale d'axe ........................................................................... m

$p_{uv}$ : pas longitudinal de la trajectoire défini dans l'espace paramétrique de la surface

$r$ : rayon de coin de l'outil ........................................................................... m

$R$ : rayon d'outil ........................................................................................ m

$s$ : abscisse curviligne de la trajectoire............................................................ m

$S$ : surface nominale

$S_1$ : surface de guidage

$S_2$ : surface d'orientation

$t$ : temps.................................................................................................. s

$Tu$ : temps d'usinage ................................................................................. s

$(u,v)$ : espace paramétrique des surfaces

$v$ : vitesse courante d'axe .......................................................................... m.s$^{-1}$

$V$ : vitesse maximale d'axe ........................................................................ m.s$^{-1}$

*Vf* : vitesse d'avance outil/pièce.................................................................................m.s$^{-1}$

*Xm,Ym,Zm* : position des axes de translation dans le repère associé à la machine.m

*Xpr,Ypr,Zpr* : positions du point piloté de l'outil dans le repère pièce....................m

$\delta$ : distance du positionnement outil calculé au plan parallèle................................m

$\theta_t$ : angle d'inclinaison de l'axe de l'outil................................................................rad

$\theta_n$ : angle de pivotement de l'axe de l'outil ............................................................rad

1. Introduction

Le processus de réalisation des pièces est composé de différentes activités constituant une chaîne numérique : définition d'un modèle de référence, construction d'un modèle de fabrication basé sur ce modèle référence, transformation des données, pilotage et suivi du procédé... etc. Le grand nombre de paramètres intervenant dans chacune de ces activités ainsi que les échanges d'informations entre les activités constituent des difficultés et des limites à la maîtrise du processus global pour respecter les critères de productivité et de qualité du produit fini.

Dans le cadre de l'usinage 5 axes à grande vitesse (UGV), il existe de nombreux verrous technologiques au niveau de chaque étape de la chaîne numérique FAO - post processeur - commande numérique (CN) - machine outil. Collisions, posage précis de l'outil ou encore prédiction de la géométrie usinée sont des points d'étude à intégrer à la génération des trajets d'usinage afin de garantir la conformité de la pièce vis-à-vis de la qualité requise [1]. Des défauts peuvent également apparaître selon les performances du couple machine outil/commande numérique utilisé (influence de la structure de la machine outil, gestion de ses défauts géométriques, commande par la CN...) [2][3]. Enfin, selon les typologies de pièces usinées, le procédé de coupe UGV nécessite également une certaine maîtrise des efforts de coupe, des déformations ainsi que des vibrations qui peuvent être générées [4].

Cette communication est dédiée à l'amélioration de l'usinage 5 axes UGV des pièces de formes complexes. Nous nous intéressons plus particulièrement aux activités de génération de trajectoires, post-processing et préparation de la trajectoire pour son suivi en cours d'usinage. Une fois la trajectoire outil/pièce calculée, les positions articulaires de la machine sont déterminées à l'aide de la transformation géométrique inverse. Ensuite, la commande numérique assure le suivi de cette trajectoire dans l'espace articulaire afin de respecter les consignes programmées. Cependant, le comportement réel des axes est limité par divers paramètres CN (vitesse, accélération et jerk maxi des axes, temps de cycle d'interpolation du DCN...) et par des problèmes liés à l'inversion de coordonnées (dépassements des courses, singularités, choix des configurations articulaires multiples) [5][6].

Après avoir exposé la chaîne numérique en usinage, les principales difficultés liées au 5 axes UGV sont analysées. Nous présentons ensuite un modèle de représentation des trajectoires 5 axes sous forme surfacique que nous avons développé, permettant d'intégrer des contraintes géométriques et cinématiques [7]. Ainsi, nous proposons d'intégrer ce modèle dans une structure d'optimisation des trajets 5 axes afin de respecter la qualité requise tout en maximisant la productivité.

2. Problématique de l'usinage 5 axes UGV

Parmi l'ensemble des procédés de fabrication, le choix se porte de plus en plus sur l'usinage à 5 axes UGV pour la réalisation des pièces de formes complexes dans le domaine de l'automobile et de l'aéronautique. L'objectif est alors de réaliser une pièce conforme d'un point de vue géométrique tout en respectant des critères de productivité liés au contexte UGV. Il convient donc de rappeler ce qu'est la chaîne numérique en usinage 5 axes, tout en mettant en avant les difficultés rencontrées et ce plus spécifiquement dans le cadre de l'usinage 5 axes continu dans un contexte UGV.

2.1. La chaîne numérique en 5 axes

Le processus de réalisation de pièces à surfaces complexes est organisé autour de trois activités principales : la conception en CAO, le calcul des trajectoires de l'outil en FAO et la fabrication sur MOCN.

Lors de la conception, le bureau d'étude établit les spécifications fonctionnelles des formes sous forme d'un modèle géométrique CAO à partir des contraintes fonctionnelles.

La génération de trajectoire consiste tout d'abord à calculer le chemin à suivre par le(s) point(s) piloté(s) de l'outil à partir du modèle CAO de la matrice. Le modèle FAO créé est classiquement constitué d'une séquence de points et de divers paramètres d'usinage. Cependant, les logiciels de CFAO génèrent des solutions génériques d'usinage dont le format n'est pas reconnu par la commande numérique qui interprète quant à elle une solution spécifique décrite en «code iso» (norme ISO 6983). Un post-processeur de traduction du langage est alors utilisé à l'interface FAO/commande numérique (Fig 1).

Ensuite, une fois le programme CN généré, le suivi de trajectoire, c'est à dire le contrôle en temps réel du mouvement relatif entre l'outil et la pièce est assuré par le directeur de commande numérique (DCN) de la MOCN. La surface usinée, état fini ou semi-fini, résulte du mouvement enveloppe de l'outil.

Il est important de rappeler qu'en 5 axes, les mouvements des axes de la machine dans l'espace des taches sont différents des trajectoires de l'outil décrites dans l'espace de la pièce par le logiciel de FAO. Ainsi, les trajectoires d'usinage 5 axes calculées par le post-processeur et communiquées à la commande numérique peuvent être décrites dans le repère pièce ou dans le repère machine. Dans le premier cas, le programme est indépendant de la structure de la machine qui usinera la pièce. Par contre, la commande numérique devra effectuer l'inversion de coordonnées en temps réel lors de l'usinage. Dans le second cas, un post-processeur dédié à la machine utilisée devra effectuer la transformation de coordonnées pour passer du repère pièce au repère machine en plus de la traduction de langage.

2.2. Erreurs et approximations liées à la chaîne numérique

Les activités associées à la chaîne numérique ne doivent pas être indépendantes. Elles doivent en particulier favoriser le dialogue et les échanges d'informations pour faciliter la réalisation du produit. En effet, diverses erreurs ou approximations présentes dans le processus de réalisation sont sources d'écarts entre les contraintes fonctionnelles associées au produit et le produit réalisé. Ces erreurs peuvent être classées dans deux catégories : les erreurs générées au sein de chaque activité et les erreurs dues aux difficultés de communication entre activités.

En conception, certaines limites sont imposées par les modeleurs géométriques actuels qui ne donnent pas une grande souplesse pour la définition des formes gauches.

En génération de trajectoire, des erreurs peuvent apparaître lors du calcul du positionnement de l'outil sur la surface. Elles résultent soit d'un positionnement peu précis de l'outil sur la surface à usiner, soit de collisions entre l'outil et la surface. Il est ainsi nécessaire de faire appel à des outils de simulation de trajectoires afin de valider les trajets générés. Cette validation reste cependant à cette étape purement géométrique.

En ce qui concerne l'activité de réalisation de la pièce sur la machine outil, les erreurs engendrées sont de diverses natures. Il faut ici séparer l'activité principale en sous activités : commande et de pilotage des axes pour le suivi de trajectoires et sous activités associées à la dynamique de l'usinage (coupe UGV, déformation en cours d'usinage, vibration des parties mécaniques, ….). L'identification des écarts est ici plus complexe. En général, les sous activités sont envisagées séparément.

De plus, à chaque activité correspond un modèle qui permet de décrire et/ou représenter le produit : des contraintes géométriques à l'étape initiale de conception, une géométrie surfacique pour la conception et une séquence de points et de paramètres pour génération de trajectoires. Ainsi, chaque activité possède un modèle différent qui ne comprend pas la même quantité et le même niveau d'informations sur le produit. Ceci explique les difficultés de communication et d'échange entre ces activités, les erreurs, les manques d'information qui apparaissent. Pour que la pièce soit conforme au cahier des charges d'un point de vue géométrique, les écarts au modèle nominal doivent être contenus. Ainsi, les diverses erreurs du processus interviennent dans ce critère d'acceptation de la pièce. Cependant, toutes les erreurs n'influent pas de la même façon, ou avec la même importance sur la géométrie de la pièce usinée.

Le tableau 1 liste de manière non exhaustive les principaux défauts observés, classés selon deux catégories : écarts au niveau de la géométrie, manques de productivité. Ce classement permet de montrer les défauts influents vis-à-vis de l'objectif de réalisation d'une pièce géométriquement conforme tout en respectant des critères de productivité. Pour chaque défaut, le phénomène physique entrant en jeu et la cause sont cités. Enfin, les paramètres influents et modifiables sont également recensés.

2.3. Les spécificités de l'usinage 5 axes UGV

Parmi ces phénomènes, causes et paramètres, certains sont valables que l'on soit en usinage conventionnel ou UGV, 3 axes ou 5 axes, mais certains sont spécifiques ou liés à l'UGV 5 axes. Ce sont soit des composants technologiques spécifiques UGV (DCN, broches, moteurs couples...), soit des techniques d'usinage adaptées à la grande vitesse (stratégies d'usinage...) soit des paramètres prenant des valeurs particulières ou ayant une grande influence sur le comportement de la machine.

D'après le tableau, nous retiendrons les paramètres récurrents, en les considérant comme étant les plus influents pour l'UGV 5 axes :

   - la section de copeau ;

   - le montage d'usinage (position - orientation pièce) ;

   - la stratégie d'usinage (trajectoire, format d'interpolation...) ;

   - le comportement - réglage du couple MOCN.

L'étude que nous présentons dans la suite s'attache plus particulièrement à explorer l'influence du comportement du couple CN/ machine-outil sur le suivi de trajectoire d'usinage 5 axes. Les comportements sont étudiés sur le centre d'usinage du LURPA, une machine Mikron UCP710 à cinématique de type RRTTT, équipée d'une commande numérique Siemens 840D.

De façon générale, en usinage 5 axes, le suivi de la trajectoire de l'outil est réalisé par déplacements coordonnés des 5 axes de la machine. Les commandes appliquées sur les axes sont calculées en temps réel par le directeur de commande numérique (DCN). Les calculs sont effectués avec les valeurs qui sont rentrées dans les paramètres 5 axes de la CN. Ainsi, la géométrie de la machine est identifiée dans la CN, et elle est considérée comme parfaite.

Comme nous l'avons souligné précédemment, les mouvements des axes de la machine dans l'espace des taches sont différents des trajectoires de l'outil décrites dans l'espace de la pièce par le logiciel de FAO. La présence d'un post-processeur est donc essentielle. Les logiciels de transformation de coordonnées intégrés dans les commandes numériques actuelles permettent d'obtenir des résultats très satisfaisants [8] mais leurs algorithmes de fonctionnement ne sont pas accessibles. C'est pourquoi nous avons développé notre propre post-processeur assurant l'inversion de coordonnées dédié à la machine utilisée [9].

Le comportement «cinématique» du couple CN – machine outil est conditionné voire limité par les caractéristiques machine, les caractéristiques du DCN et les fonctions du DCN. En particulier, la vitesse d'avance atteinte par l'outil lors d'un suivi de trajectoire ne correspond pas à la vitesse programmée et ce en raison des phénomènes majeurs suivants :

- les limites des performances cinématiques des axes :

Les axes possèdent certaines caractéristiques mécaniques. Afin de préserver les composants mécaniques, la CN réduit les valeurs maximales qui peuvent être atteinte par des limites logicielles. Chaque axe possède ainsi une vitesse maximale, une accélération maximale et un jerk maximal. Le comportement des axes est géré de manière à ce que ces contraintes soient toujours respectées, ce qui est un facteur limitant, en particulier pour la coordination des axes lors du suivi de la trajectoire théorique. Les axes sont coordonnés pour suivre la trajectoire articulaire voulue et l'on constate en particulier que les axes rotatifs sont souvent limitants en terme de vitesse et accélération.

- les caractéristiques du DCN :

Dans la phase de préparation de la trajectoire, le temps de cycle d'interpolation limite la vitesse programmée sur les petits segments pour assurer le calcul des consignes de position.

- les fonctions du DCN :

Dans un contexte UGV, il est conseillé de choisir un mode de passage des discontinuités dans l'espace articulaire de la machine (saut de vitesse) qui permet d'arrondir la trajectoire au voisinage de la discontinuité. Ainsi, on adapte la vitesse en fin de bloc, ce qui peut créer des ralentissements.

La combinaison de l'usinage multi-axes dans un contexte UGV impose par ailleurs d'utiliser certaines fonctions du DCN dont le choix résulte d'un compromis entre productivité et respect de la qualité [10]. Ainsi, on choisira un mode de pilotage de type Soft pour lequel le profil d'accélération est trapézoïdal. Ce mode permet de préserver la mécanique tout en maintenant les erreurs à la trajectoire confinées. Cependant, c'est un mode de pilotage plus lent que le mode de pilotage à profil d'accélération rectangulaire. De la même manière, afin de maîtriser le comportement des axes, la gestion des écarts est réalisée dans l'espace articulaire, c'est-à-dire axe par axe (G642).

Ainsi, les phénomènes présentés plus haut agissent comme des facteurs limitants, que ce soit d'un point de vue géométrique (erreurs conduisant à des écarts géométriques) que d'un point de vue cinématique (limitation de la vitesse d'avance). De ce fait, nous proposons de présenter ces facteurs sous forme de contraintes géométriques et cinématiques. Nous présentons dans la suite le modèle de représentation des trajectoires 5 axes sous forme surfacique que nous avons développé, permettant de prendre en compte ces contraintes. Ce modèle est ensuite intégré dans une structure d'optimisation des trajets 5 axes en vue de respecter la qualité requise tout en maximisant la productivité.

3. Optimisation à l'aide d'un modèle d'usinage surfacique

Le modèle d'usinage surfacique a été défini par Tournier et al dans [11]. Dans le cadre de l'usinage 5 axes en bout, le modèle d'usinage est défini par deux surfaces biparamétrées (Fig. 2) : la surface de guidage ($S_1$) et la surface d'orientation ($S_2$). La surface de guidage assure le posage de l'outil sur la surface, quelle que soit son orientation. L'orientation de l'axe de l'outil est gérée indépendamment par la deuxième surface ($S_2$).

Le modèle d'usinage sous forme surfacique, proche de la notion de la peau de la pièce permet d'assurer une continuité dans deux directions. De plus, de par sa définition, il autorise un découplage entre la position de l'outil et son orientation par rapport à la pièce.

La génération de trajectoire revient alors à choisir selon la stratégie retenue un parcours particulier sur les surfaces de guidage et d'orientation. La surface de guidage assure la conformité de la pièce. L'intégration des contraintes dues à l'usinage 5 axes UGV est envisagée au travers de l'optimisation de la surface d'orientation.

3.1. Structure d'optimisation

Dans le cadre d'optimisation des trajets, nous proposons d'intégrer le modèle d'usinage surfacique dans l'architecture plus globale présentée en figure 3.

La première étape consiste à calculer une instance du modèle surfacique à partir d'un modèle de référence de la pièce, instance à partir de laquelle la trajectoire 5 axes est calculée selon la stratégie pré-définie. Après calcul de la trajectoire sur chaque axe dans l'espace articulaire par le post-processeur, le comportement cinématique de la machine outil est analysé par un modèle prédictif du couple CN – machine outil [12].

Si la qualité géométrique n'est pas respectée, une modification de la planification de la trajectoire est nécessaire pour rendre la pièce usinée conforme. Une fois la conformité de la pièce assurée, nous proposons deux voies pour l'optimisation de la productivité : soit une modification du posage de la pièce dans l'espace de travail soit une modification de la surface de guidage et/ou de la surface d'orientation. La modification de la surface de guidage permet de « lisser » les sollicitations ou discontinuités situées dans l'espace articulaire de la machine tout en respectant les écarts géométriques. La modification de la surface d'orientation quant à elle permet d'améliorer le suivi de trajectoires en intégrant les contraintes cinématiques.

Dans la partie suivante, nous présentons la deuxième voie d'optimisation de la productivité au travers de la modification de la surface d'orientation.

3.2. Construction du modèle d'usinage et génération de trajectoires 5 axes

La construction du modèle d'usinage et la génération de trajectoires s'articulent autour des trois points suivants :

1 - Construction de la surface de guidage à partir du modèle de référence ;

2 - Définition d'une instance de la surface d'orientation à partir d'un mode de balayage et d'un mode de gestion de l'orientation de l'axe de l'outil ;

3 - Calcul des positions et des orientations de l'outil en fonction des paramètres de la stratégie d'usinage retenue.

Le choix de la géométrie de l'outil est réalisé à partir d'une analyse géométrique du modèle de référence (courbures, accessibilité, forme à générer...). La surface de guidage est obtenue par décalage selon la normale de la surface à usiner (*S*) d'une valeur égale au rayon de coin, *r*, de l'outil (Fig 2) :

$$S_1(u,v) = S(u,v) + r.n(u,v) \tag{1}$$

La surface d'orientation est définie par le calcul d'une surface offset généralisée en fonction des modes de balayage et de gestion de l'orientation de l'axe outil. Nous avons retenu un mode de balayage par plans parallèles dans le cas de l'interpolation linéaire et une orientation de l'axe de l'outil constante par rapport à la surface :

$$S_2(u,v) = S(u,v) + r.n(u,v) + (R-r).v(u,v) \qquad (2)$$

L'orientation peut être optimisée pour intégrer les contraintes cinématiques tout en assurant des trajectoires hors collision.

Le calcul du positionnement de l'outil en fonction des paramètres retenus est réalisé de manière discrète. La première étape consiste à définir la position et le nombre de plans de guidage dans l'espace cartésien en fonction des paramètres de la stratégie. La distance entre les plans, $D$, est calculée à partir d'une première estimation de la hauteur de crête générée entre deux passes. Elle est réévaluée si nécessaire lors de l'optimisation.

La seconde étape est dédiée au calcul des positions de l'outil dans l'espace paramétrique de la surface de guidage afin d'annuler les erreurs de posage. Les intersections entre les plans et les frontières du carreau sont tout d'abord calculées par cheminement sur le long des bords de l'espace paramétrique. Elles définissent pour chaque passe, le premier et le dernier point de posage outil ($S_1(u_0,v_0)$, $S_1(u_f,v_f)$). Les positions successives de l'outil sont calculées de proche en proche : à partir de la position ($u_i,v_i$) et d'un pas longitudinal ($p_{uv}$), la position suivante ($u^*_{i+1},v^*_{i+1}$) est estimée dans l'espace paramétrique par la formule ci-dessous :

$$\begin{vmatrix} u^*_{i+1} \\ v^*_{i+1} \end{vmatrix} = \begin{vmatrix} u_i \\ v_i \end{vmatrix} + \frac{p_{uv}}{\sqrt{(u_f - u_i)^2 + (v_f - v_i)^2}} \cdot \begin{vmatrix} u_f - u_i \\ v_f - v_i \end{vmatrix} \qquad (3)$$

La distance $\delta^*$ entre la position estimée $S_I(u^*_{i+1}, v^*_{i+1})$ et le plan considéré est alors calculée. Un raffinement de type Newton Raphson est utilisé pour réduire la valeur $\delta^*$ de façon à projeter le point sur le plan (Fig. 4).

L'erreur de corde générée par l'interpolation linéaire entre deux positions successives est évaluée à partir de la courbure locale de la surface dans la direction d'usinage. Si l'écart est supérieur à la valeur autorisée, la démarche détaillée ci-dessus est réitérée en diminuant le pas d'avance $p_{uv}$ dans l'espace paramétrique de la surface.

3.3. Calculs en vue d'une optimisation cinématique

A l'aide du post-processeur dédié à la Mikron UCP710, nous effectuons la transformation géométrique inverse pour obtenir les consignes de position des 5 axes de la machine (*Xm,Ym,Zm,A,C*) à partir des positionnements de l'outil calculés (*Xpr,Ypr,Zpr,I,J,K*). Une fois les consignes articulaires obtenues, elles sont introduites dans le modèle prédictif de comportement cinématique afin d'obtenir l'évolution des positions $p(t)$, vitesses $v(t)$, accélérations $a(t)$ et jerks $j$ de chacun des axes. Entre deux configurations articulaires interpolées par la CN, les déplacements en temps réel des axes au travers de leur pilotage en jerk sont soumis à des contraintes liées entre autres aux performances des moteurs [12] :

$$\begin{cases} p(t) = p_0 + v(s) \cdot t + \frac{1}{2} a(s) \cdot t^2 + \frac{1}{6} j \cdot t^3 \\ |v(s)| \leq V(s)_{axe\ limitant} \\ |a(s)| \leq A(s)_{axe\ limitant} \\ j = J(s)_{axe\ limitant} \end{cases} \quad (4)$$

Le calcul des profils cinématiques effectué dans un espace adimensionné nous permet de mettre en évidence quels sont les axes limitants vis à vis de la vitesse, de l'accélération et du jerk le long de chaque trajectoire $C(s)= S_1(u(s),v(s))$ définie sur la surface de guidage.

En ce qui concerne le problème d'optimisation, une approche simplifiée consiste à minimiser le temps de parcours de la trajectoire $C(s)$ soumis aux contraintes cinématiques définies par (4) :

$$\text{minimiser} \quad Tu = \int_0^L \frac{1}{Vf(s)} \, ds \tag{5}$$

où $Vf$ représente la vitesse relative outil/pièce recalculée à partir du modèle géométrique direct, et qui s'exprime ainsi comme une fonction implicite des angles d'inclinaison et de pivotement $\theta_t(s)$ et $\theta_n(s)$. L'optimisation consiste alors à déterminer les lois d'évolution des angles d'orientation $\theta_t(s)$ et $\theta_n(s)$ le long des trajets permettant de maximiser les performances cinématiques. Le problème ainsi posé ne peut se résoudre de manière explicite.

Notre objectif dans cet article étant de valider la démarche d'optimisation s'appuyant sur un modèle surfacique de trajectoire, nous proposons de déterminer des orientations améliorant les performances cinématiques par déformations locales de la surface d'orientation.

4. Application

L'application est effectuée sur un paraboloïde hyperbolique (surface réglée de degré 2) modélisé sous forme d'un carreau NURBS. La stratégie d'usinage retenue consiste en un guidage selon les règles à 45° avec une distance entre passes fixe de 8 mm. L'outil torique utilisé possède un grand rayon de 5 mm et un rayon de coin de 1.5 mm. L'angle d'inclinaison initial vaut 1° et l'angle de pivotement est nul. Dans une première approche, la surface d'orientation est représentée de manière polyédrique à partir d'une discrétisation de la surface de guidage (Fig. 5).

Une fois les trajectoires générées, le modèle prédictif fait apparaître une saturation en vitesse sur l'axe C, la vitesse de rotation maximale de l'axe C étant de 20 tr/min. L'axe C est donc l'axe limitant vis-à-vis du suivi de trajectoire comme nous pouvons le constater sur la figure 6 gauche partie ombrée. Nous proposons donc de modifier les orientations de l'axe outil dans une des zones où l'axe C sature afin d'atteindre au mieux la vitesse d'avance programmée. La pièce étant symétrique, nous ne nous intéressons qu'à la moitié gauche de la pièce. Nous avons analysé l'évolution des vitesses des axes rotatifs avec différents angles d'inclinaison. A priori, il existe des zones pour lesquelles un angle d'inclinaison de 5° permet de moins saturer, tout en conservant une inclinaison de 1° là où il n'y a pas de solution. Aussi, nous avons introduit des modifications locales par déformation de la surface d'orientation (Fig. 6 milieu) en imposant à quelques points une inclinaison de 5°.

Grâce à la continuité de la surface d'orientation, toute une zone est déformée et n'importe quelle position de l'outil sur la passe considérée est modifiée. Dans la direction transversale à la passe, cette déformation a également généré une modification de l'angle d'inclinaison sur la passe précédente et sur la passe suivante. On observe sur la figure 6 droite que les saturations sur l'axe C sont diminuées, ce qui valide la pertinence de la démarche d'optimisation à partir de la déformation de la surface d'orientation.

5. Conclusion

Dans le cadre de la génération de trajectoires d'usinage en fraisage 5 axes à grande vitesse, le choix de la stratégie d'usinage et de l'orientation de l'axe de l'outil doit permettre de favoriser le suivi de la trajectoire. Dans ce sens nous proposons de générer les trajectoires d'usinage à l'aide d'un modèle surfacique assurant le découplage des contraintes géométriques et cinématiques. Ce modèle est basé sur la donnée d'une surface de guidage assurant le respect des spécifications géométriques couplé à la donnée d'une surface d'orientation permettant d'optimiser les orientations de l'axe de l'outil afin de garantir un meilleur suivi cinématique de la trajectoire dans l'espace articulaire.

Par ailleurs, nous avons développé un modèle afin de prévoir la vitesse effective de l'outil par rapport à la pièce le long de la trajectoire en fonction des caractéristiques du DCN et des caractéristiques des axes de la machine. Associé au post-processeur, ce modèle prédictif permet d'avoir une meilleure image des positions et vitesses requises sur chacun des axes de la machine pour garantir qualité géométrique et productivité. Nous avons montré qu'une analyse a posteriori couplée à une modification de la surface d'orientation permet en particulier de limiter les saturations des axes les plus limitants lors du suivi de trajectoire.

L'étape suivante porte sur l'écriture du problème d'optimisation global à partir des équations (4) et (5). La méthode de résolution pourra s'appuyer sur la donnée de lois d'évolution des angles d'inclinaison et de pivotement afin d'obtenir les orientations de l'axe de l'outil optimisées vis-à-vis du suivi de trajectoire.

| Impact sur : | Phénomène entrant en jeu | Causes possibles | Paramètres influents |
|---|---|---|---|
| Géométrie | Déformation outil | Efforts de coupe | Section copeau – Matériau outil<br>Attachement outil (technologie)<br>Géométrie de l'outil (diamètre, géométrie de l'extrémité, longueur) |
| Géométrie | Déformation pièce | Efforts de coupe | Section copeau<br>Montage d'usinage (rigidité)<br>Matériau de la pièce<br>Mode d'obtention du brut<br>Trajectoire d'usinage (séquencement) |
| Géométrie | Vibrations<br>Dynamique | Efforts de coupe<br>Rigidité – Structure – Modes propres des ensembles (pièce/porte pièce) et (outil/broche/axes/bâti) | Trajectoire d'usinage (moins accidentée, modifier la stratégie d'usinage)<br>Montage d'usinage<br>Outil (matériau, attachement, longueur) |
| Géométrie | Mouvement relatif outil/pièce (enveloppe du mouvement outil = pièce) | Génération de trajectoire au sens calcul | Valeurs des paramètres d'usinage<br>Format d'interpolation<br>Méthode de calcul du posage outil |
| Géométrie | Mouvement relatif outil/pièce (enveloppe du mouvement outil = pièce) | Défauts géométriques des liaisons géométriques de la machine et des règles de mesure | |
| Productivité<br>Géométrie | Mouvement de l'outil dans le repère machine ou/et mouvement relatif outil/pièce non désirés (singularité géométrique) | Architecture machine<br>Génération de trajectoire (géométrie)<br>DCN (transformation géométrique) | Montage d'usinage (posage + orientation pièce)<br>Programme CN (points)<br>Mode de pilotage du DCN<br>Architecture machine |
| Géométrie | Dilatation thermique | Frottements – échauffements | Mise en chauffe – Stabilité thermique |
| Productivité<br>Géométrie | Vitesse d'avance réelle limitée | Saturation des moteurs (surtout les moteurs rotatifs) | Posage pièce (sollicitation des axes différente – rotation – translation) |
| Productivité<br>Géométrie | Vitesse d'avance réelle limitée | DCN (temps de cycle faible) | Format d'interpolation<br>Trajectoire (longueur des segments) |
| Productivité<br>Géométrie | Ralentissements de la vitesse d'avance | Inertie des axes (accélération – jerk maxi) | Trajectoire (changements de direction, courbure, mode de balayage, longueur des segments)<br>Réglage des asservissements CN |

Tab. 1 : Défauts influents sur la qualité géométrique et la productivité

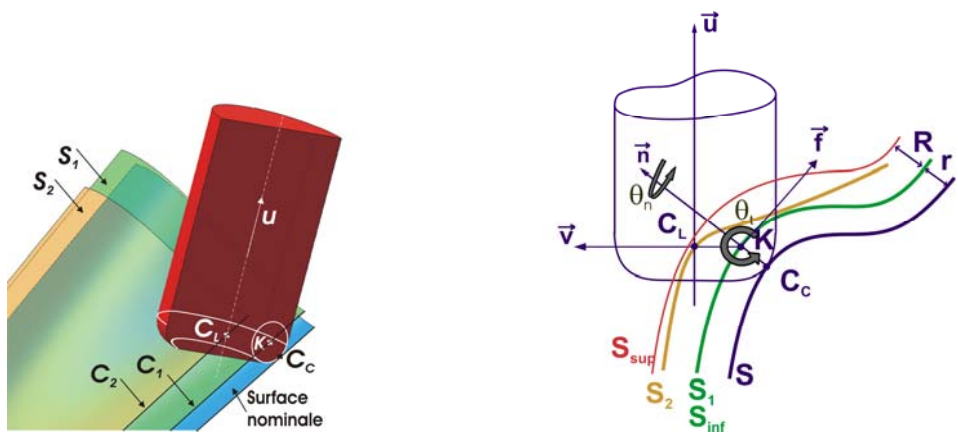

Fig. 2 : Modèle d'usinage surfacique en 5 axes

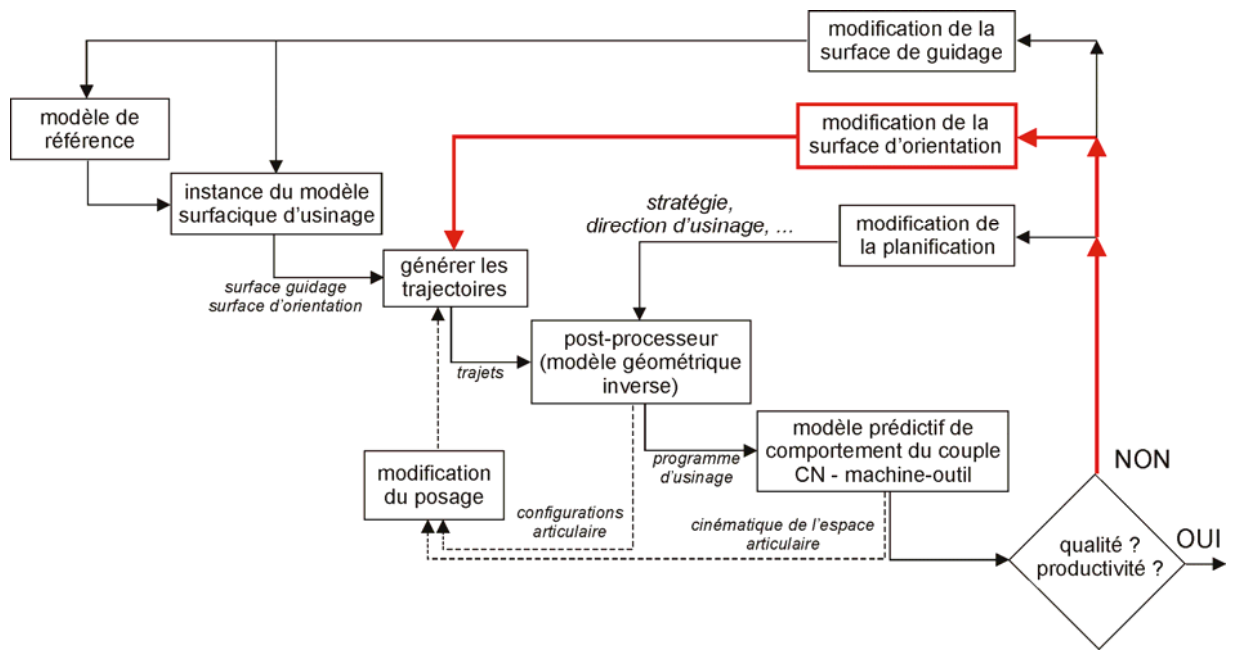

Fig. 3 : Structure d'optimisation globale

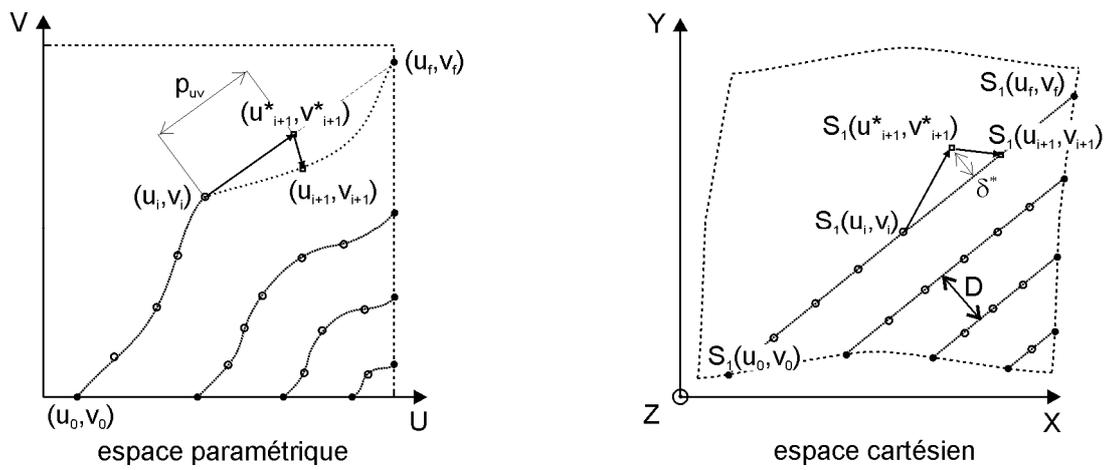

Fig. 4 : Calcul des positions de l'outil

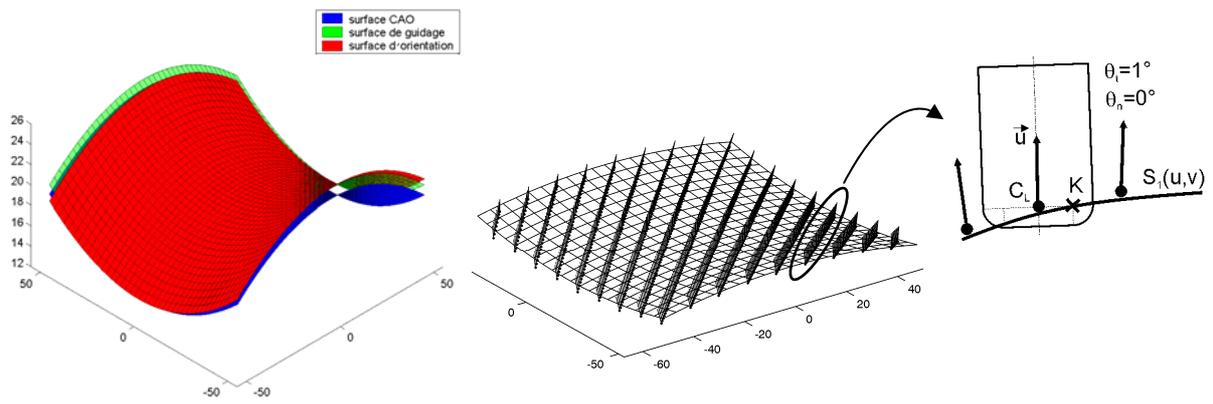

Fig. 5 : Positions et orientations outil calculées sur l'instance du modèle

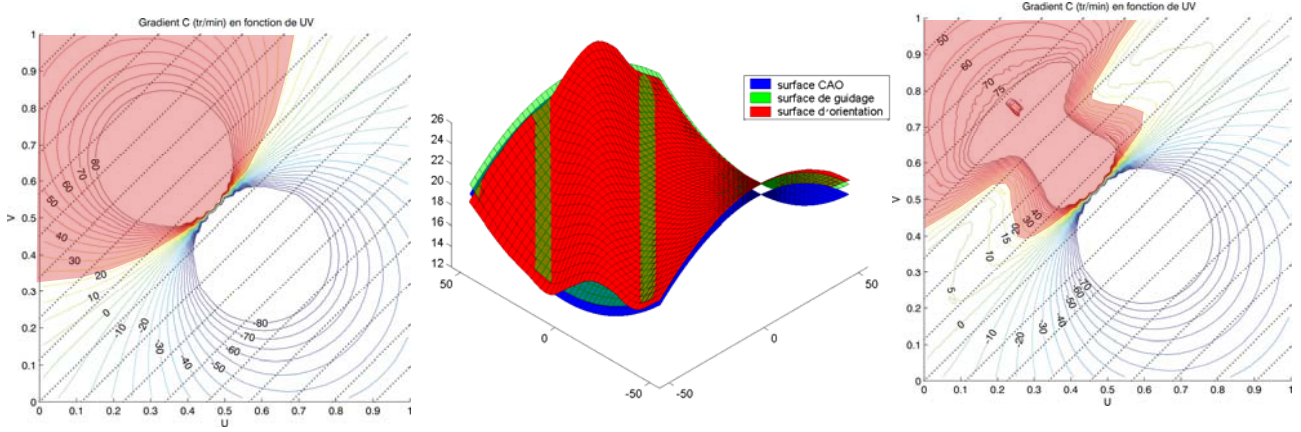
Fig. 6 : Surface d'orientation déformée, influence sur la saturation de l'axe C